\documentclass[a4,ugly]{ion}
\usepackage{graphicx}
\usepackage{amsmath}
\usepackage{url}
\usepackage{caption}
\usepackage{dblfloatfix}
\usepackage{bm}
\usepackage{subcaption}
\usepackage{tabularx}
\usepackage{amssymb}
\makeatletter
\def\hlinewd#1{%
\noalign{\ifnum0=`}\fi\hrule \@height #1 %
\futurelet\reserved@a\@xhline}
\makeatother
\title{Computationally Efficient Unscented Kalman Filtering Techniques for Launch Vehicle Navigation using a Space-borne GPS Receiver}

\author{Sanat K. Biswas, \textit{ACSER, UNSW Australia}\\ 
Li Qiao, \textit{UNSW Australia}\\
Andrew G. Dempster, \textit{ACSER, UNSW Australia}}
\begin{document}
\maketitle
\section*{Biography}
\bigskip
\textbf{Sanat Biswas} is a PhD student in the School of Electrical Engineering and Telecommunications at the University of New South Wales (UNSW). He received BE in Instrumentation and Electronics from Jadavpur University and M. Tech in Aerospace Engineering from Indian Institute of Technology Bombay. Sanat is currently associated with Australian Centre for Space Engineering Research (ACSER) and Satellite Navigation and Positioning (SNAP) Laboratory. His research focus is non-linear estimation techniques for on-board space vehicle navigation using GNSS receiver. He has been awarded the Emerging Space Leaders Grant 2014 by the International Astronautical Federation.\\
\\
\textbf{Dr. Li Qiao} is a Research Associate in the School of Engineering and Information Technology at the University of New South Wales (UNSW), Canberra. She joined UNSW as a visiting PhD student from 2009 to 2010, and obtained her PhD in Guidance, Navigation and Control at Nanjing University of Aeronautics and Astronautics in 2011. Her research interests are satellite orbit modelling, satellite autonomous navigation and integrated navigation.\\
\\
\textbf{Professor Andrew Dempster} is Director of the Australian Centre for Space Engineering Research (ACSER) in the Sch-ool of Electrical Engineering and Telecommunications at the University of New South Wales (UNSW). He has a BE and MEngSc from UNSW and a PhD from the University of Cambridge in efficient circuits for signal processing arithmetic. He was system engineer and project manager for the first GPS receiver developed in Australia in the late 80s and has been involved in satellite navigation ever since. His current research interests are in satellite navigation receiver design and signal processing, areas where he has six patents, and new location technologies. He is leading the development of space engineering research at ACSER.
\begin{abstract}
\bigskip
The Extended Kalman Filter (EKF) is a well established technique for position and velocity estimation. However, the performance of the EKF degrades considerably in highly non-linear system applications as it requires local linearisation in its prediction stage. The Unscented Kalman Filter (UKF) was developed to address the non-linearity in the system by deterministic sampling. The UKF provides better estimation accuracy than the EKF for highly non-linear systems. However, the UKF requires multiple propagations of sampled state vectors in the measurement interval, which results in higher processing time than for the EKF. This paper proposes an application of two newly developed UKF variants in launch vehicle navigation. These two algorithms, called the Single Propagation Unscented Kalman Filter (SP-
UKF) and the Extrapolated Single Propagation Unscented Kalman Filter (ESPUKF), reduce the processing time of the original UKF significantly and provide estimation accuracies comparable to the UKF. The estimation performance of the SPUKF and the ESPUKF is demonstrated using Falcon 9 V1.1 launch vehicle in CRS-5 mission scenario. The launch vehicle trajectory for the mission is generated using publicly available mission parameters. A SPIRENT GNSS simulator is used to generate the received GPS signal on the trajectory. Pseudo-range observations are used in the EKF, UKF, SPUKF and the ESPUKF separately and the estimation accuracies are compared. The results show that the estimation errors of the SPUKF and the ESPUKF are 15.44\% and 10.52\% higher than the UKF respectively. The processing time reduces by 83\% for the SPUKF and 69.14\% for the ESPUKF compared to the UKF.
\end{abstract}
\section{Introduction}
\bigskip
Accurate navigation of a launch vehicle is crucial for every space mission. Launch vehicle position and velocity information is required for insertion of spacecraft into their orbits and for range safety. Navigation of launch vehicles involves extensive real-time ground based radar tracking and communications \cite{Whiteman2005}. However, with advancements in Global Navigation Satellite Systems (GNSS), integration of GNSS measurements with existing navigation techniques for launch vehicles has become conspicuous. Global Positioning System (GPS) measurements are combined with the traditional dead-reckoning navigation measurements and ground based radar measurements to obtain accurate navigation data in real time \cite{farrell2001carrier, Ailneni2013, Minor1998}. However, due to the highly non-linear nature of launch vehicle dynamics, it is a challenging problem to estimate position and velocity with minimal uncertainty using GPS/GNSS measurements. 

The Extended Kalman Filter (EKF) is a widely used estimation technique to combine the knowledge of the dynamics of the user vehicle motion with the GNSS/GPS measurements for robust and more accurate position and velocity solutions. In the prediction stage of the EKF, the non-linear system model is linearized to compute the \textit{a priori} error covariance matrix. This linearisation results in a degraded state estimation for highly non-linear and high dynamic system \cite{Julier2000}. In order to solve this problem, Julier and Uhlmann suggested a deterministic sampling technique to compute the \textit{a priori} error covariance matrix to avoid local linearisation of a non-linear system \cite{Julier1997,Julier1998,Julier2000,Julier2004}. This approach is widely known as the Unscented Kalman Filter (UKF) and it appears to be an effective option for non-linear estimation. However, the UKF used for continuous dynamic systems requires more processing time than the EKF \cite{Sarkka2007}. Therefore, the UKF is often not the preferred technique for real-time estimation application due to its computationally expensive nature. This paper presents the performance of two new computationally efficient variants of the UKF called the Single Propagation Unscented Kalman Filter (SPUKF) and the Extrapolated Single Propagation Unscented Kalman Filter (ESPUKF) in a launch vehicle position and velocity estimation scenario with GPS measurements.

In the prediction stage of the UKF several sigma points are calculated from the \textit{a posteriori} mean state vector and the error covariance at an epoch. These sigma points are separately propagated using a numerical integration technique to the next epoch and the \textit{a priori} mean state vector and the error covariance are calculated. In the SPUKF only the \textit{a posteriori} mean state vector is propagated to the next epoch and the deviation of the other sigma points from the \textit{a posteriori} mean is utilized to calculate the other sigma points at the new epoch. The calculation involves evaluation of the Jacobian matrix for the non-linear function corresponding to the launch vehicle motion and Taylor Series approximation \cite{TACBiswas2015}. The \textit{a priori} mean state vector and the error covariance are calculated from these sigma points.

The processing time of the SPUKF can be reduced by 90\% as compared to the UKF \cite{Biswas2015}. However, this first-order approximation results in an estimation error of the order of the second-order Taylor Series terms. In the ESPUKF the second-order terms from the estimation are eliminated using the Richardson Extrapolation technique \cite{TACBiswas2015}. In previous contributions the authors developed a simulation setup for testing and verification of an KF based satellite position estimation algorithm using multi-GNSS measurements \cite{ASRCBiswas2014,IACBiswas2014}. A similar simulation experiment was carried out to demonstrate the SPUKF and the ESPUKF performance in a scenario involving launch vehicle navigation using GPS. In this work, the SpaceX Falcon 9 V1.1 launch vehicle used in the Commercial Resupply Service (CRS)-5 was selected as the test scenario. The launch vehicle trajectory was simulated using publicly available launch vehicle and mission specific data \cite{SpaceX2009, CRS5}. A SPIRENT GNSS simulator was used to simulate the GPS measurements for the user launch vehicle. The simulator provided pseudo-range and carrier-range data were processed using the EKF, UKF, SPUKF and the ESPUKF respectively for comparison.
\section{Kalman Filter for Non-linear Systems}
\bigskip
The Kalman Filter (KF) is an optimal estimation technique for linear systems \cite{Kalman1960}. However, this optimal estimation technique can not be used in most of the practical applications because of the non-linearity in physical systems. The EKF, a sub-optimal variant of the KF is the most popular estimation algorithm for non-linear estimation problems. In the prediction stage of the EKF the \textit{a posteriori} state vector at an epoch is propagated using the non-linear differential equation of the system to calculate the \textit{a poriori} mean state vector at the new epoch. However the error covariance is propagated by linearising the system equation. The error due to the linearisation is compensated by adding an arbitrary fudge factor in the process noise matrix $\bm Q$ to stabilize the solution. Sometimes a high value of the elements in the fudge factor is chosen for solution stability. This results in more dependency on the measurement during the correction stage and in turn the solution becomes more susceptible to spurious measurement error. To address the system non-linearity more accurately, Julier and Uhlmann suggested a deterministic sampling technique referred as the Unscented Transform (UT) in the prediction stage of the estimation process to calculate the mean state vector and the error covariance matrix \cite{Julier2000}. A Kalman Filter with the UT in the prediction stage is referred as a UKF. In the UT, the state vector is augmented with the elements of the process noise vector and the error covariance matrix is augmented with the process noise matrix \cite{Julier1997}. From the augmented state vector and the error covariance matrix the sampled state vectors are computed deterministically at the current epoch. If the number of elements in the augmented state vector is $n$ then the number of samples required is $2n+1$. These sampled state vectors are called the sigma points \cite{Julier2000, Julier2004}. To compute the \textit{a priori} mean state vector at the next epoch, these $2n+1$ sigma points are propagated separately and then a weighted average is taken \cite{Julier2000}. Subsequently the error covariance at the next epoch is calculated from the weighted mean state vector and the sigma points \cite{Julier2000}. Similarly, measurements are computed using the measurement model for each of the sigma points and weighted average is considered as the predicted measurement. The UT approach of state vector prediction results in higher computation time compared to the EKF due to the requirement for multiple state propagations in a single time step and this makes implementation of the UKF challenging in a system with limited computation power for real-time computation.

 In our previous work \cite{TACBiswas2015}, two new approaches to the state prediction were suggested within the Unscented Kalman Filter framework to improve computational efficiency. In the first method, only one augmented state vector containing the \textit{a posteriori} state elements is propagated to the next epoch. The other $2n$ sigma points at the next epoch are approximated using 1st order Taylor series approximation. This approximation requires computation of the Jacobian matrix of the system and the matrix exponential of the Jacobian. As the number of state propagations in every time step is reduced to one, the computation time reduces significantly \cite{TACBiswas2015}. The UKF with the new state prediction technique is called the SPUKF. However, due to the first-order approximation of the sigma points at the new epoch, the error in the estimation comprises the second-order Taylor Series terms \cite{TACBiswas2015}. To eliminate the second order terms from the state estimation, the Richardson Extrapolation is utilized in the ESPUKF \cite{TACBiswas2015}. In this method, sigma points at the new epoch are computed using the Richardson Extrapolation technique, in which the second-order Taylor Series terms are included in the sigma point approximation. The ESPUKF proves to be more accurate than the SPUKF with a minor increase in the processing time. The ESPUKF can deliver an estimation accuracy similar to the UKF with significant reduction in processing time.
\section{Mathematical Model of Launch Vehicle Dynamics}
\begin{figure}[h!]%
\includegraphics[width=\columnwidth]{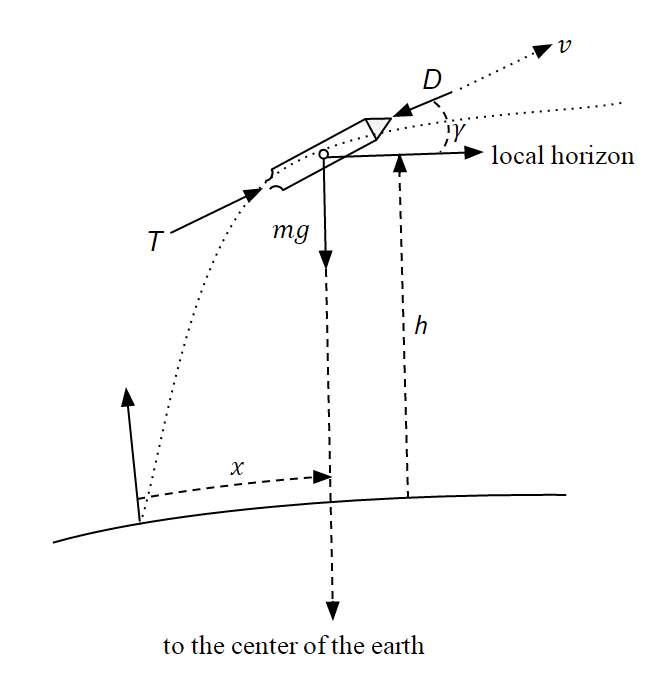}%
\caption{Launch vehicle trajectory}%
\label{fig:LV_diag}%
\end{figure}

A typical launch vehicle trajectory is shown in the Figure \ref{fig:LV_diag}, where $x$ is the down-range distance, $h$ is the altitude, $v$ is the speed and $\gamma$ is the flight path angle of the launch vehicle under consideration. State elements considered for the estimation are $x, h, v, \gamma$, aerodynamic coefficient $C$, mass $m$, GPS receiver clock bias $b$ and receiver clock bias rate $\dot b$.
The state vector is defined as 
\begin{equation}
\bm X = \left(\begin{array}{c}
	x\\
	h\\
	v\\
	\gamma\\
	m\\
	C\\
	b\\
	\dot b
\end{array}\right)
\label{eq:state_vec}
\end{equation}
The system model can be expressed as \cite{Curtis2010}
\begin{equation}
\bm{\dot X} = \left(\begin{array}{l}
								\frac{R_E}{R_E + h}v\cos{\gamma}\\
								v\sin{\gamma}\\
								\frac{T}{m} - \frac{D}{m} - g\sin{\gamma}\\
								-\frac{1}{v}\left(g - \frac{v^2}{R_E + h}\right)\cos{\gamma}\\
								-\dot{m}_e\\
								0\\
								\dot b\\
								0
								\end{array}\right) + \bm{\nu}(t)
\label{eq:LV_model}
\end{equation}
where $\dot{m}_e$ is the mass flow rate at the exhaust nozzle, $T$ is the thrust provided by the engine of the current stage, $D$ is the aerodynamic drag, $g$ is the gravitational acceleration and $R_E$ is the local radius of the earth. $\bm{\nu}(t)$ is a $8\times 1$ process noise vector. Ideally $T$ remains constant till the burnout time of a stage and changes to a different value depending on the engine characteristics of the next stage. $D$ depends on the frontal area of the launch vehicle.
 
\subsection{Mission specific parameters}
To demonstrate the performance of the SPUKF and the ESPUKF in a launch vehicle navigation problem, the CRS-5 mission scenario was selected. In the CRS 5 mission a Falcon 9 V1.1 launch vehicle was used. The launch vehicle delivered a Dragon cargo spacecraft in space to resupply the International Space Station (ISS) . The mission and launch vehicle specific parameters for the scenario is provided in Table \ref{tab:LV_param} \cite{SpaceX2009, CRS5}.
\begin{table}[h!]%
\centering
\caption{Mission and Launch Vehicle Specific Parameters}
\begin{tabular}{ll}
\hline
Mission parameters &\\ \hline
Payload & 2317 kg\\
Dragon spacecraft mass & 4200 kg\\
Orbit perigee & 410 km\\
Orbit apogee & 418 km\\ \hline
Stage 1 & \\ \hline
Inert Mass & 23,100 kg\\
Propellant Mass & 395,700 kg\\
Engine & $9\times$ Merlin 1D\\
Thrust & 5886 kN \\
Specific Impulse & 282 s \\
Burnout Time & 187 s \\ \hline
\end{tabular}
\label{tab:LV_param}
\end{table} 
\begin{table}[h!]%
\centering
\begin{tabular}{ll}
\hline
Stage 2 & \\ \hline
Inert Mass & 3,900 kg\\
Propellant Mass & 92,670 kg\\
Engine & $1\times$ Merlin 1D Vac\\
Thrust & 801 kN \\
Specific Impulse & 340 s \\
Burnout Time & 386 s \\ \hline
\end{tabular}
\label{tab:LV_param1}
\end{table} 

\section{GPS Measurement model}
Pseudo-range and carrier range measurements of the GPS are modelled as \cite{kaplan2005understanding,Misra2006} :
\begin{align}
\rho_{i}(t) = & r_i(t) + c[\delta t_u(t) - \delta t_i(t - \tau)]\nonumber \\
						 & + {\textit I}(t) + {\textit T}(t) + \epsilon_{\rho}(t)
\label{eq:Prange}
\end{align}

\begin{align}
\Phi_{i}(t) = & r_i(t) + c[\delta t_u(t) - \delta t_i(t - \tau)]\nonumber \\
						 & + {\textit I}_{\phi}(t) + {\textit T}_{\phi}(t) + \lambda N + \epsilon_{\Phi}(t)
\label{eq:Crange}
\end{align}
where\\
\\ \begin{tabular}{cl}
 $i$ & is GNSS satellite index\\
 $\rho_i$ & is pseudo-range from the launch vehicle\\
          & to the navigation satellite $i$\\
 $\Phi_i$ & is carrier-range from the launch vehicle\\
          & to the navigation satellite $i$\\
 $r_i$ & is geometric distance from the launch vehicle\\
       & to the navigation satellite $i$\\
 $\delta t$& is receiver clock bias\\
 $\delta t_i$& is clock bias of the navigation satellite\\
 $\tau$ & is signal transmission time\\
 $c$ & is velocity of light\\
 ${\textit I}(t)$& is ionospheric error for pseudo-range\\
 ${\textit T}(t)$& is tropospheric error for pseudo-range\\
\end{tabular}\\
\begin{tabular}{cl}
 ${\textit I}_{\Phi}(t)$& is ionospheric error for carrier range\\
 ${\textit T}_{\Phi}(t)$& is tropospheric error for carrier range\\
 $\lambda$ & is the wavelength of the carrier signal\\
 $N$ & is integer ambiguity\\
 $\epsilon_{\rho}(t)$& is random noise in pseudo-range measurement\\
 $\epsilon_{\Phi}(t)$& is random noise in carrier-range measurement   
\end{tabular}\\

The tropospheric error is calculated using the Saastamoinen model \cite{Misra2006} and the ionospheric error is eliminated from the pseudo-range using the GRAPHIC technique \cite{jones1993environmental} after resolving the integer ambiguity from the carrier-range. 
\section{Implementation of Unscented Filters}
In unscented filtering, the evolution of the process noise statistics over time is addressed by augmenting the state vector with the process noise terms \cite{Julier1997}. The augmented state vector is
\begin{equation}
\bm{X}_a(t) = \begin{bmatrix}
											  \bm{X}(t)\\
												\bm{W}(t)
											\end{bmatrix}	
\label{eq:augmented}
\end{equation}
In the UKF, the sigma points are calculated from \cite{Julier2000}
\begin{equation}
\bm{X}^+_a(t) = \begin{bmatrix}
													\bm{X}^+(t)\\
													\bm{0}_{8\times 1}
													\end{bmatrix}
\label{eq:augmented_0}
\end{equation}
\begin{equation}
\bm{P}_a(t) = \left[\begin{array}{ccc}
\bm{P}(t) & \bm{P}_{XW}(t)\\
\bm{P}_{XW}(t) & \bm{Q}(t)
\end{array}\right]
\label{eq:augmented_cov}
\end{equation} 
Here, $\bm{X}^+(t)$ and $\bm{X}^+_a(t)$ are the \textit{a posteriori} state vector and the augmented state vector respectively at epoch $t$. The augmentation terms are zero because the process noise distribution is considered as zero mean Gaussian. $\bm{P}(t)$ and $\bm{P}_a(t)$ are the error covariance and the augmented error covariance matrix respectively. $\bm{P}_{XW}(t)$ is the cross covariance of $\bm X$ and $\bm W$. $\bm Q(t) = \bm E[\bm W \bm W^T]$ is the process noise covariance matrix. The dimension of the augmented state vector is 16. Therefore, a total of 33 sigma points must be propagated to the next epoch to predict the weighted \textit{a priori} mean state vector and the error covariance. The sigma points and the corresponding weights are
\begin{align}
\bm X_0(t) &= \bm{X}^+_a(t)\\
\bm X_i(t) &= \bm{X}^+_a(t) + \bm{\Delta X}_i, (i = 1,2,3...32)\\
W_0 &= \frac{\kappa}{n+\kappa}\\
W_i &= \frac{1}{2(n+\kappa)}, (i = 1,2,3...32)\label{eq:sigma_point}
\end{align}
and
\begin{center}
\begin{tabular}{lcl}
$\Delta{\bm{X}}_i$ &= $(\sqrt{(n+\kappa){\bm{P}}}_a)_i$ & for $i = 1,2,3....16$\\
$\Delta{\bm{X}}_i$ &= $-(\sqrt{(n+\kappa){\bm{P}}}_a)_i$ & for $i = 17,2,3....32$\\
\end{tabular}
\end{center}
$(\sqrt{(n+\kappa)\bm{P}_a})_i$ is the $i$th column of the matrix $\sqrt{(n+\kappa)\bm{P}_a}$.
$\kappa$ is a parameter and generally it is selected in such a way that $(n+\kappa) = 3$\cite{Julier2000}. Corresponding to all the 33 propagated sigma points the measurement vectors are computed using the measurement equation \ref{eq:Prange} and the weighted mean of them is considered to be the predicted measurement vector. The measurement error covariance and the cross covariance between measurement vector and the state vector is computed using the predicted mean state and measurement vector, the predicted sigma points and the corresponding measurement vectors \cite{Julier2000}. Then the conditional mean state vector and the error covariance is computed using the Kalman Filter equations \cite{Julier2000}.

\begin{figure*}%
\includegraphics[scale = .4]{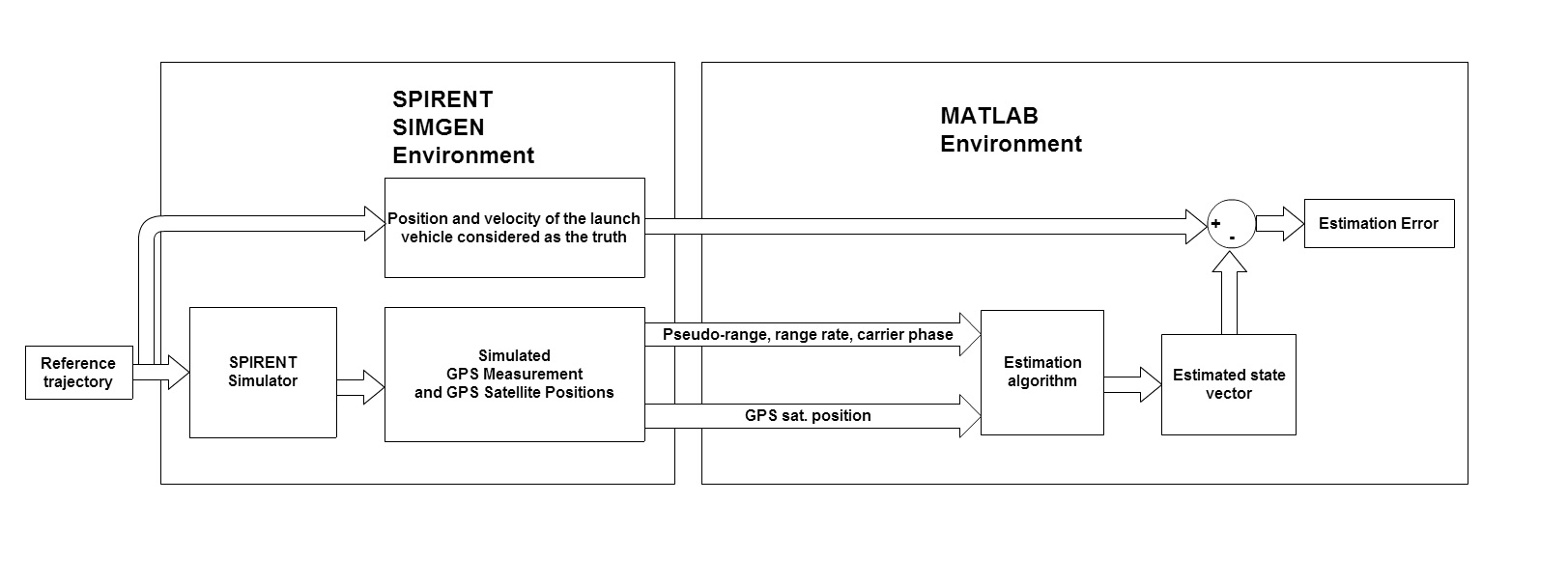}%
\caption{Simulation Setup with SPIRENT and MATLAB}%
\label{fig:sim}%
\end{figure*} 

\subsection{Single propagation Unscented Kalman Filter}
In the SPUKF, only $\bm{X}_0(t)$ is propagated to the next epoch. The other sigma points are not propagated. To calculate the sigma points at the next epoch $t + \delta t$, the following equation is utilized \cite{TACBiswas2015}
\begin{equation}
\bm{X}_i^-(t+\delta t) = \bm{X}_0^-(t+\delta t) + e^{\bm{\mathcal{J}}\delta t}{\Delta \bm{X}_i}
\label{eq:sigma_approx}
\end{equation}
Here $\bm{X}_0^-(t+\delta t)$ propagated augmented state vector at $t+\delta t$ and
\begin{align}
\bm{\mathcal{J}} & = \left.\frac{\partial \dot{\bm X}_a}{\partial \bm X_a}\right|_{\bm{X}^+_a(t)}\nonumber\\
&= \left[\begin{array}{cc}
\left.\frac{\partial \dot{\bm X}}{\partial \bm X}\right|_{\bm X^+(t)} & \bm{0}_{8\times 8}\\
\bm{0}_{8\times 8} & \bm{0}_{8\times 8}																																	
\end{array}\right]
\label{eq:jacobi}
\end{align}
After calculation of all the sigma points the standard weighted mean and covariance calculation method of the UT \cite{Julier2000} is used to compute the \textit{a priori} mean state vector and the error covariance matrix. The correction stage of the SPUKF is the same as the UKF.

\subsection{Extrapolated Single propagation Unscented Kalman Filter}
In the ESPUKF, the sigma points are computed using the following equations \cite{TACBiswas2015}:
\begin{align}
 N_1\bm{ (\Delta X_i)} &= \bm{X}_0^-(t+\delta t) + e^{\bm{\mathcal{J}}\delta t}{\Delta \bm{X}_i}
\label{eq:N1}\\
N_2\bm{(\Delta X_i)} &= \bm{X}_0^-(t+\delta t) + e^{\bm{\mathcal{J}}\delta t}\frac{\Delta \bm{X}_i}{2}\nonumber\\
											&	+ e^{\bm{\mathcal{J}'}\delta t}\frac{\Delta \bm{X}_i}{2}\label{eq:N2}\\
\bm{X}_i^-(t+\delta t) &= 2N_2\bm{(\Delta X_i)} - N_1\bm{ (\Delta X_i)}
\label{eq:richardson}
\end{align}
Here, 
\begin{align}
\bm{\mathcal{J}'} & = \left.\frac{\partial \dot{\bm X}_a}{\partial \bm X_a}\right|_{\bm{X}^+_a(t) + \frac{\Delta \bm{X}_i}{2}}\nonumber
\end{align}
Computation of sigma points using equation \ref{eq:richardson} results in inclusion of the second-order Taylor series terms in the approximation \cite{TACBiswas2015}. The rest of the calculation procedure in the ESPUKF is the same as for the SPUKF.
\section{Simulation}
To demonstrate the performance of the SPUKF and the ESPUKF for a launch vehicle navigation scenario, a reference trajectory was generated using equation \ref{eq:LV_model} and table \ref{tab:LV_param} for Falcon 9 V1.1 launch vehicle. The simulation setup is shown in Figure \ref{fig:sim}. This reference trajectory was used as the input to the SPIRENT GNSS simulator. The GPS measurements for the trajectory and the GPS satellite positions generated by the simulator are used in different estimation algorithms separately in a MATLAB environment and the performance were compared. The pseudo-range and range rates are used as measurements and random noise is incorporated artificially in the measurements to simulate measurement noise. The state and the error covariance for the filter initialization are:
\begin{align}
\widehat{\bm{X}}(0) &= \left(\begin{array}{l}
0\ m\\
0\ m\\
5.6543\ ms^{-1}\\
1.5708\ rad\\
5.20\times10^5\ kg\\
0.5010\\
400\ m\\
2\	ms^2
\end{array}\right)\nonumber
\label{eq:filt_state}
\end{align}

\begin{align}
\bm{P}(0) &= diag\left(\begin{array}{l}
	1\\
	1\\
	0.01\\
	10^{-6}\\
	9\\
	0.01\\
	9\times10^4\\
	25
\end{array}\right)\nonumber
\end{align}

The process noise covariance matrix is considered as
\begin{align}
\bm{Q} &= 10^{-30}\bm{I}_{9\times 9}\nonumber
\end{align}
Due to high dynamics of the launch vehicle, it may not be possible for a GPS receiver to acquire GPS signals through all the available channels through out the trajectory. To examine the performance of various estimation algorithms during limited availability of the acquired signals, multiple simulations were performed by restricting the number of channels to 4, 6, 8 and 10. 

\subsection{Results}
\begin{figure}[h!]%
\includegraphics[width=\columnwidth]{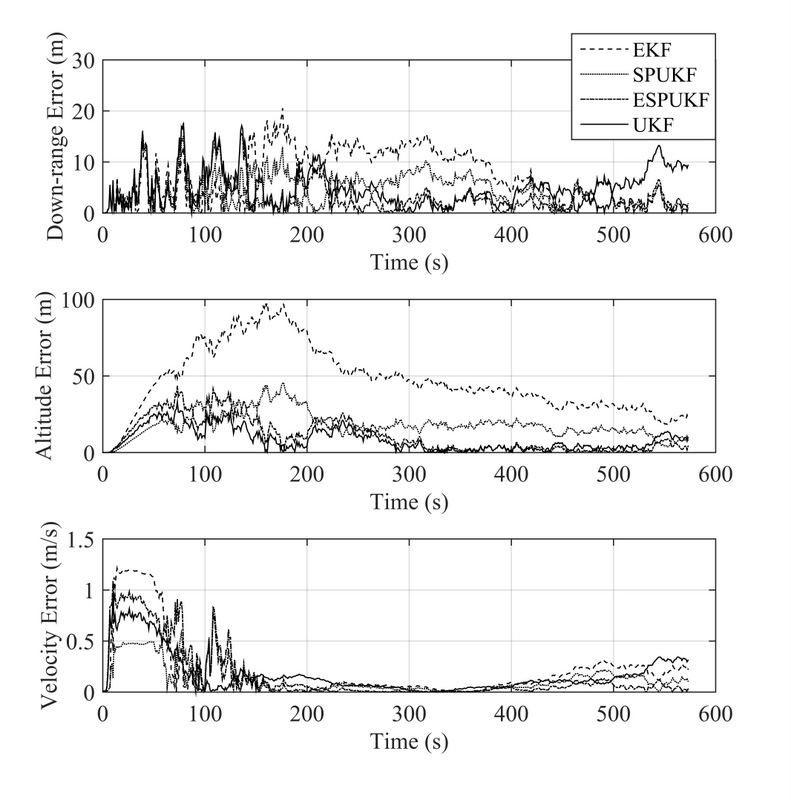}%
\caption{Estimation error using different algorithms}%
\label{fig:er_comp}%
\end{figure}

In Figure \ref{fig:er_comp} the down-range, altitude and speed estimation errors for the EKF, UKF, SPUKF and the ESPUKF are shown for 6 channel observation. It can be observed that the estimation error for the SPUKF, ESPUKF and the UKF is less than that of the EKF. 
\begin{figure}[h!]%
\includegraphics[width=\columnwidth]{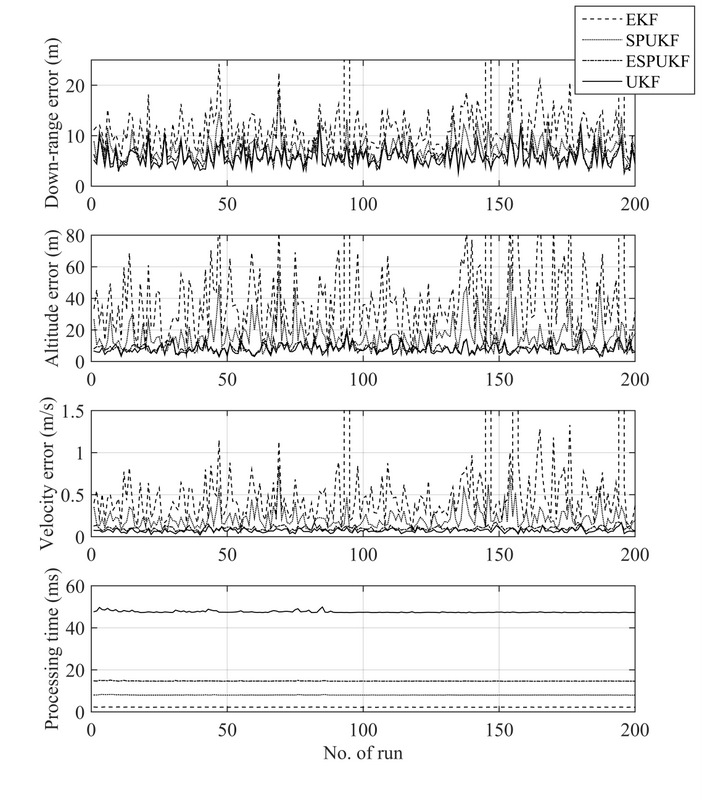}%
\caption{Average error for different filters using 4 channels}%
\label{fig:NS4}%
\end{figure}
\begin{figure}[h!]%
\includegraphics[width=\columnwidth]{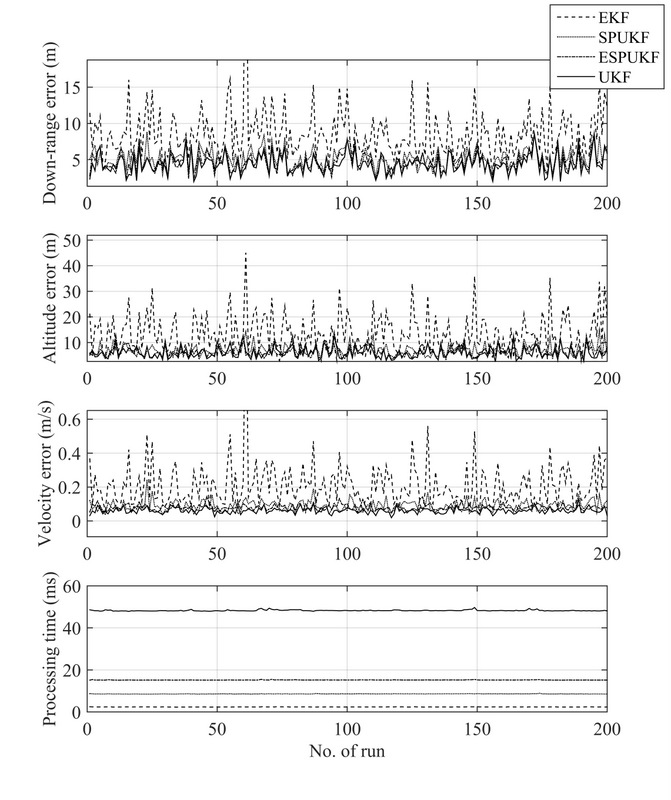}%
\caption{Average error for different filters using 6 channels}%
\label{fig:NS6}%
\end{figure}
\begin{figure}[h!]%
\includegraphics[width=\columnwidth]{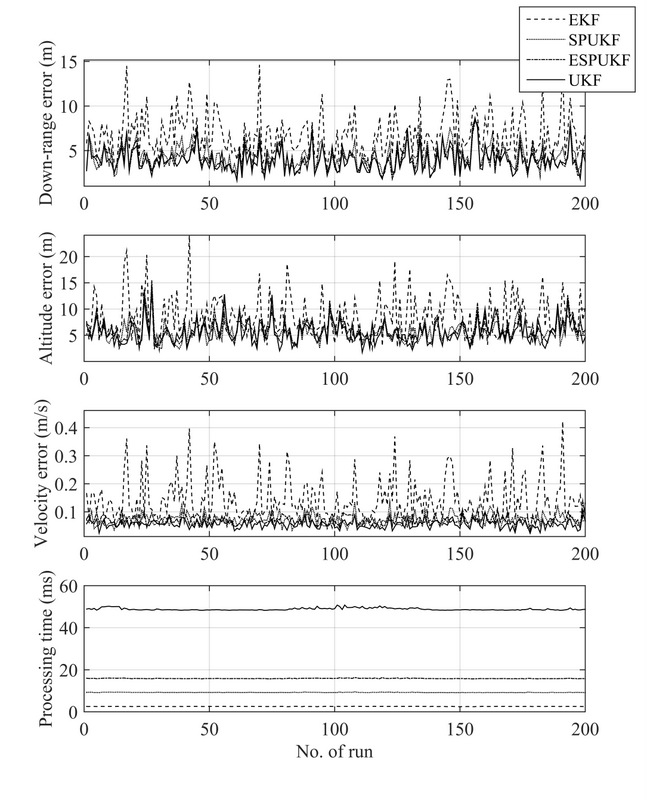}%
\caption{Average error for different filters using 8 channels}%
\label{fig:NS8}%
\end{figure}
\begin{figure}[h!]%
\includegraphics[width=\columnwidth]{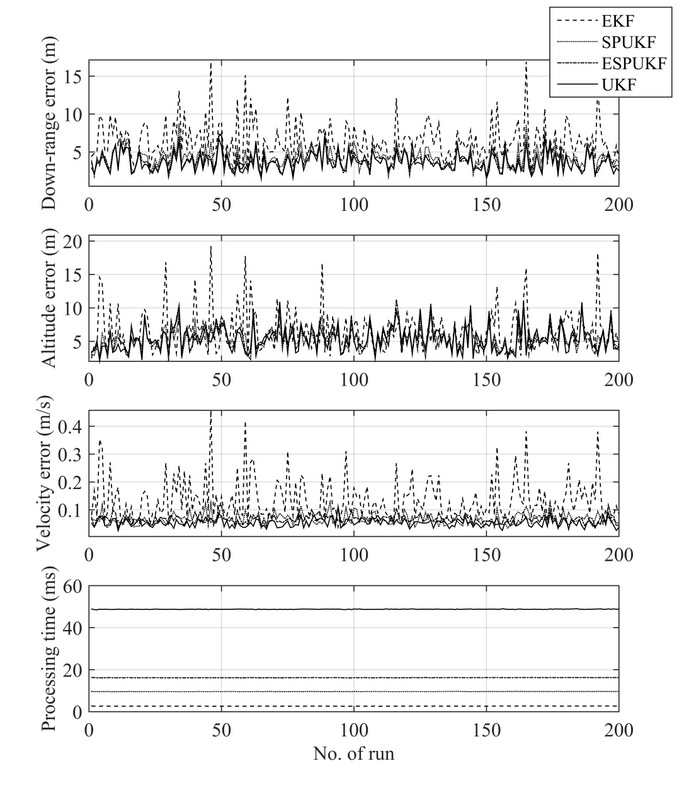}%
\caption{Average error for different filters using 10 channels}%
\label{fig:NS10}%
\end{figure}
\begin{figure}[h!]%
\includegraphics[width=\columnwidth]{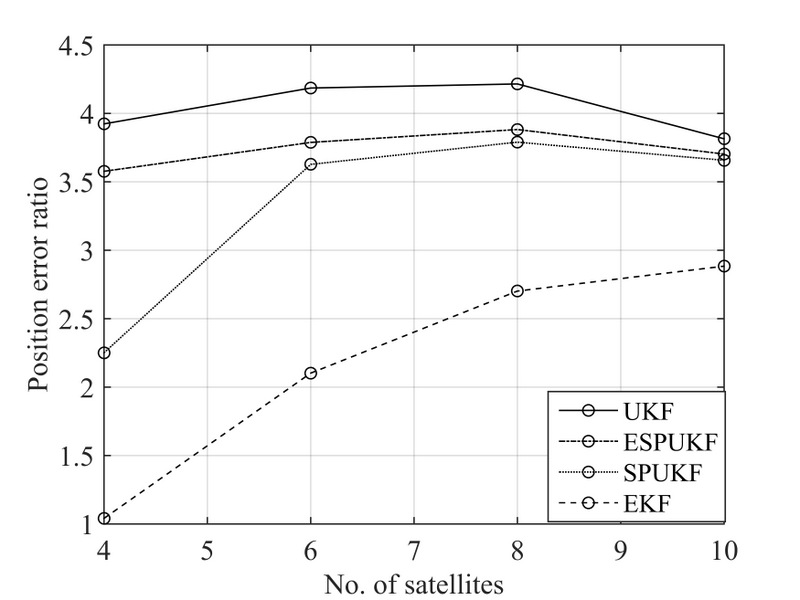}%
\caption{Position error ratio for different number of GPS satellites used}%
\label{fig:KF_DOP}%
\end{figure}
\begin{figure}[h!]%
\includegraphics[width=\columnwidth]{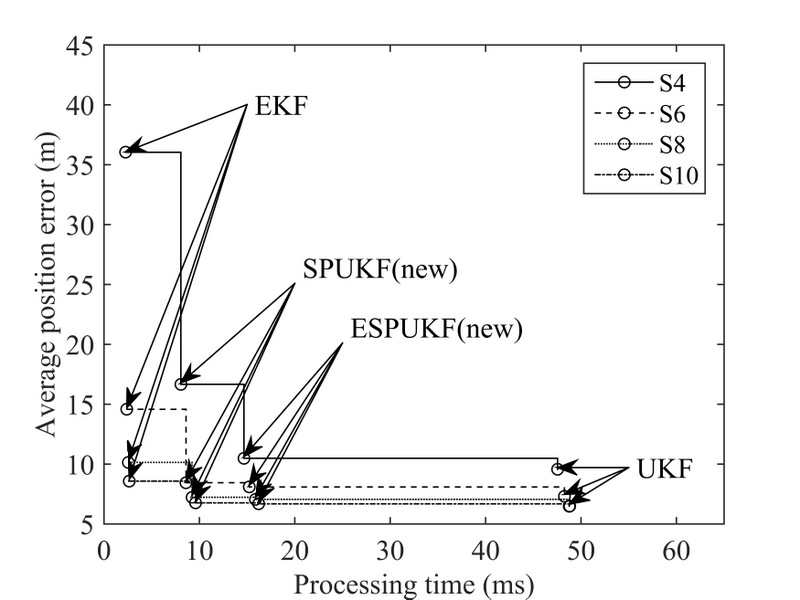}%
\caption{Processing time vs. estimation error}%
\label{fig:KF_PT}%
\end{figure}

For consistency checking, 200 simulations were performed for 4, 6, 8 and 10 channels separately. For each simulation, random noise was generated and added to the measurements before performing the estimation. In Figures \ref{fig:NS4}, \ref{fig:NS6}, \ref{fig:NS8} and \ref{fig:NS10}, the time average of estimation errors for down-range, altitude and velocity using different filters are shown. It is observed that, with an increase in the number of channels i.e. number of satellites used for navigation, the difference in estimation errors for the EKF and other Unscented Filters decreases.
To further understand the trend, position error ratio for different estimation algorithms vs. number of satellites used is plotted in Figure \ref{fig:KF_DOP}. The position error ratio for each estimation technique was defined as:
\begin{equation}
\textrm{Position error ratio} = \frac{PDOP\times \sigma_R}{\textrm{median position error}}
\label{eq:PER}
\end{equation}
Here, $PDOP$ is the Position Dilution of Precision and $\sigma_R$ is the standard deviation of the pseudo-range noise. Median position error of the 200 simulations is considered to avoid the effect of the outliers of the EKF estimation errors (the EKF diverges in those runs) which can be observed in Figures \ref{fig:NS4}, \ref{fig:NS6}, \ref{fig:NS8} and \ref{fig:NS10}. Figure \ref{fig:KF_DOP} implies that, in the Kalman Filter framework, the ratio between the position error and the standard deviation of the pseudo-range noise is not $PDOP$ and this ratio is different for different filters. It should be also noted that, the position error ratio increases with increase in number of satellites and then decreases.   

\begin{table}%
\caption{Performance of different filters with different number of GPS signals}
\begin{tabular}{l c c}
\hlinewd{1pt}
\multicolumn{2}{l}{No. of GPS observations: 4}& \\ \hline
					& Position error (m) & Processing time (ms)\\ \hline
 EKF & 36.04 & 2.26\\
 SPUKF & 16.66 & 8.06\\
 ESPUKF & 10.49 & 14.67 \\
 UKF & 9.56 & 47.54\\ \hlinewd{1pt}
\multicolumn{2}{l}{No. of GPS observations: 6}& \\ \hline
					& Position error (m) & Processing time (ms)\\ \hline
 EKF & 14.58 & 2.39\\
 SPUKF & 8.45 & 8.60\\
 ESPUKF & 8.09 & 15.22 \\
 UKF & 7.32 & 48.26\\ \hlinewd{1pt}
\multicolumn{2}{l}{No. of GPS observations: 8}& \\ \hline
					& Position error (m) & Processing time (ms)\\ \hline
 EKF & 10.14 & 2.55\\
 SPUKF & 7.23 & 9.23\\
 ESPUKF & 7.06 & 15.86 \\
 UKF & 6.50 & 48.80\\ \hlinewd{1pt}
\multicolumn{2}{l}{No. of GPS observations: 10}& \\ \hline
					& Position error (m) & Processing time (ms)\\ \hline
 EKF & 8.58 & 2.66\\
 SPUKF & 6.76 & 9.60\\
 ESPUKF & 6.68 & 16.20 \\
 UKF & 6.48 & 48.76\\ \hlinewd{1pt}
\end{tabular}
\label{tab:prf_sum}
\end{table}

In Figure \ref{fig:KF_PT} the median of time average position error for 200 simulations is plotted with average processing time required per time step for the EKF, UKF, SPUKF and the ESPUKF for different number of satellites used. It can be discerned from the figure that the SPUKF and the ESPUKF provide better estimation accuracy than the EKF and at the same time, both the new filters require significantly less processing time than the UKF. In Table \ref{tab:prf_sum} the estimation performance and the processing time for the EKF, SPUKF, ESPUKF and the UKF are provided for various observation cases.

It is observed that, the processing time does not increase greatly with the increase in the number of satellites used. This is because, the state propagation time is very high compared to the measurement prediction time in all the Kalman Filters and the state propagation time does not change with the number of satellites used.   
\section{Conclusion}
\bigskip
In this paper an application of two new variants of the Unscented Kalman Filter called the SPUKF and the ESPUKF is proposed for Launch Vehicle navigation using a GPS receiver. The results confirm that:
\begin{enumerate}
	\item Unscented Filtering can provide more accurate launch vehicle navigation solution than the EKF.
	\item The SPUKF and the ESPUKF reduce the launch vehicle position and velocity estimation time significantly compared to the conventional UKF. The data provided in Table \ref{tab:prf_sum} indicate that, the processing time of the SPUKF and the ESPUKF can be reduced by 83\% and 69.14\% respectively, than that of the UKF. the estimation errors of the SPUKF and the ESPUKF are 15.44\% and 10.52\% higher than the UKF respectively for six observations. The errors become similar for higher observations. 
\end{enumerate}
The results also show that, the ratio of the position error and the standard deviation of the measurement noise is different than $PDOP$ in the Kalman Filter framework. Also, the ratio is different for different filters and varies with the number of satellites selected. In our future work, a new factor will be  introduced to establish the relation of the position error, the PDOP and the measurement error standard deviation for different types of Kalman Filter. To create a more realistic scenario, a UNSW-Kea receiver, which is capable of acquiring signal during high acceleration and jerk, will be included in the launch vehicle simulation setup in future and the estimation performance of the SPUKF and the ESPUKF will be verified.
\bibliographystyle{ieeetr}
\bibliography{biblog}

\begin{thebibliography}{10}

\bibitem{Whiteman2005}
D.~E. Whiteman, L.~M. Valencia, and J.~C. Simpson, ``{Space-based range safety
  and future space range applications},'' in {\em International Association for
  the Advancement of Space Safety Conference}, no.~1, 2005.

\bibitem{farrell2001carrier}
J.~L. Farrell, ``Carrier phase processing without integers,'' in {\em
  Proceedings of the 57th Annual Meeting of the Institute of Navigation},
  pp.~423--428, 2001.

\bibitem{Ailneni2013}
S.~Ailneni, S.~K. Kashyap, and S.~K. N, ``{INS/GPS Fusion for Navigation of
  Unmanned Aerial Vehicles},'' in {\em ICIUS}, no.~3, 2013.

\bibitem{Minor1998}
R.~Minor and D.~Rowe, ``{Utilization of GPS/MEMS-IMU for measurement of
  dynamics for range testing of missiles and rockets},'' in {\em Position
  Location and Navigation Symposium, IEEE 1998}, pp.~602--607, 1998.

\bibitem{Julier2000}
S.~Julier, J.~Uhlmann, and H.~Durrant-Whyte, ``{A new method for the nonlinear
  transformation of means and covariances in filters and estimators},'' {\em
  IEEE Transactions on Automatic Control}, vol.~45, pp.~477--482, mar 2000.

\bibitem{Julier1997}
S.~Julier and J.~Uhlmann, ``{A New Extension of the Kalman Filter to Nonlinear
  Systems},'' in {\em SPIE}, vol.~3068, pp.~182--193, Orlando, FL, 1997.

\bibitem{Julier1998}
S.~J. Julier, ``{A Skewed Approach to Filtering},'' in {\em Proceedings of
  SPIE--the international society for optical engineering}, vol.~3373,
  pp.~271--282, 1998.

\bibitem{Julier2004}
S.~Julier and J.~Uhlmann, ``{Unscented Filtering and Nonlinear Estimation},''
  {\em Proceedings of the IEEE}, vol.~92, pp.~401--422, mar 2004.

\bibitem{Sarkka2007}
S.~S{\"{a}}rkk{\"{a}}, ``{On unscented Kalman filtering for state estimation of
  continuous-time nonlinear systems},'' {\em IEEE Transactions on Automatic
  Control}, vol.~52, no.~9, pp.~1631--1641, 2007.

\bibitem{TACBiswas2015}
S.~K. Biswas, L.~Qiao, and A.~G. Dempster, ``{A Novel \textit{a priori} State
  Computation Strategy for Unscented Kalman Filter to Improve Computational
  Efficiency},'' {\em IEEE transactions on Automatic Control}, 2016.

\bibitem{Biswas2015}
S.~K. Biswas, L.~Qiao, and A.~Dempster, ``{Application of a Fast Unscented
  Kalman Filtering Method to Satellite Position Estimation using a Space-borne
  Multi-GNSS Receiver},'' in {\em ION GNSS+}, 2015.

\bibitem{ASRCBiswas2014}
S.~Biswas, L.~Qiao, and A.~Dempster, ``{Space-borne GNSS based orbit
  determination using a SPIRENT GNSS simulator},'' in {\em 15th Australian
  Space Research Conference, Adelaide, Australia}, 2014.

\bibitem{IACBiswas2014}
S.~Biswas, L.~Qiao, and A.~Dempster, ``{Real-Time on-Board Satellite Navigation
  Using Gps and Galileo Measurements},'' in {\em 65th International
  Astronautical Congress, Torronto, Canada}, pp.~2--6, 2014.

\bibitem{SpaceX2009}
``{Falcon 9 Launch Vehicle Payload User's Guide},'' tech. rep., 2015.

\bibitem{CRS5}
``{SpaceX CRS-5 Fifth Commercial Resupply Services Flight to the International
  Space Station},'' tech. rep., NASA, 2014.

\bibitem{Kalman1960}
R.~E. Kalman, ``{A New Approach to Linear Filtering and Prediction Problems},''
  {\em Transactions of the ASME-Journal of Basic Engineering}, vol.~82,
  no.~Series D, pp.~35--45, 1960.

\bibitem{Curtis2010}
H.~D. Curtis, {\em {Orbital Mechanics for Engineering Students}}.
\newblock Butterworth-Heinemann, 2010.

\bibitem{kaplan2005understanding}
E.~D. Kaplan and C.~J. Hegarty, ``{Understanding GPS: principles and
  applications},'' in {\em Understanding GPS: principles and applications},
  pp.~237--260, Artech house, 2005.

\bibitem{Misra2006}
P.~Misra and P.~Enge, {\em {Global Positioning System : Signals , Measurements
  and Performance}}.
\newblock Massachusetts: Ganga-Jamuna Press, 2006.

\bibitem{jones1993environmental}
T.~P. Yunck, ``{Coping with the atmosphere and ionosphere in precise satellite
  and ground positioning},'' in {\em Environmental Effects on Spacecraft
  Positioning and Trajectories} (A.~V. Jones, ed.), Geophysical Monograph
  Series, Wiley, 1993.

\end{thebibliography}
\end{document}